\documentclass[a4paper,11pt]{article}
\pdfoutput=1 

\usepackage{jheppub} 

\usepackage[T1]{fontenc} 
\usepackage[textsize=footnotesize,textwidth=2.5cm]{todonotes}

\definecolor{mikadoyellow}{rgb} {0.16, 0.254, 0.6}

\usepackage{amssymb,amsmath,latexsym,bm,amsfonts}
\usepackage{graphicx}
\usepackage{longtable}
\usepackage{color,xcolor}
\usepackage{indentfirst}
\usepackage{subfigure}
 \usepackage{comment}

\newcommand{\be}{\begin{equation}}
\newcommand{\ee}{\end{equation}}
\newcommand{\bpm}{\begin{pmatrix}}
\newcommand{\epm}{\end{pmatrix}}

\newcommand{\beqn}{\begin{eqnarray}}
\newcommand{\eeqn}{\end{eqnarray}}

\newcommand{\besub}{\begin{subequations}}
\newcommand{\eesub}{\end{subequations}}
\newcommand{\bea}{\begin{eqnarray}}
\newcommand{\eea}{\end{eqnarray}}
\newcommand{\p}{\partial}

\def\p{\partial}

\newcommand{\ba}{\begin{array}}
\newcommand{\ea}{\end{array}}
\newcommand{\bi}{\begin{itemize}}
\newcommand{\ei}{\end{itemize}}

\title{ \boldmath Fine structure in holographic entanglement and entanglement contour}
\author[a,b,c,d]{Qiang Wen}

\affiliation[a]{Shing-Tung Yau Center of Southeast University, Nanjing 210096, China}

\affiliation[b]{Yau Mathematical Sciences Center, Tsinghua University, Beijing, 100084, China}

\affiliation[c]{Center of Mathematical Sciences and Applications, Harvard University, Cambridge, 02138, USA}

\affiliation[d]{Graduate School, China Academy of Engineering Physics, Beijing 100193, China}

\emailAdd{qwen@gscaep.ac.cn}

\abstract{We explore the fine structure of the holographic entanglement entropy proposal (the Ryu-Takayanagi formula) in AdS$_3$/CFT$_{2}$. With the guidance from the boundary and bulk modular flows we find a natural slicing of the entanglement wedge with the modular planes, which are codimension one bulk surfaces tangent to the modular flow everywhere. This gives an one-to-one correspondence between the points on the boundary interval $\mathcal{A}$ and the points on the Ryu-Takayanagi (RT) surface $\mathcal{E}_{\mathcal{A}}$. In the same sense an arbitrary subinterval $\mathcal{A}_2$ of $\mathcal{A}$ will correspond to a subinterval $\mathcal{E}_2$ of $\mathcal{E}_{\mathcal{A}}$. This fine correspondence indicates that the length of $\mathcal{E}_2$ captures the contribution $s_{\mathcal{A}}(\mathcal{A}_2)$ from $\mathcal{A}_2$ to the entanglement entropy $S_{\mathcal{A}}$, hence gives the contour function for entanglement entropy. Furthermore we propose that $s_{\mathcal{A}}(\mathcal{A}_2)$ in general can be written as a simple linear combination of entanglement entropies of single intervals inside $\mathcal{A}$. This proposal passes several nontrivial tests. }

\begin{document} 
\maketitle
\flushbottom
\section{Introduction}
The study of entanglement entropy, which describes the correlation structure of a quantum system, has played a central role in the study of modern theoretical physics in the last decade. Large amount of interest is stimulated by the Ryu-Takayanagi (RT) \cite{Ryu:2006bv,Ryu:2006ef} formula in the context of AdS/CFT correspondence \cite{Maldacena:1997re,Gubser:1998bc,Witten:1998qj}. More explicitly, for a static subregion $\mathcal{A}$ in the boundary CFT and a minimal surface $\mathcal{E}_{\mathcal{A}}$ in the dual AdS bulk that anchored on the boundary $\partial\mathcal{A}$ of $\mathcal{A}$, the RT formula states that the entanglement entropy of $\mathcal{A}$ is measured by the area of $\mathcal{E}_{\mathcal{A}}$ in Planck units,\begin{align}
S_{EE}=\frac{Area (\mathcal{E}_{\mathcal{A}})}{4 G}\,.
\end{align}
The covariant version of the RT formula is proposed in \cite{Hubeny:2007xt} with the minimal surface generalized to the extremal surface.

The are two main strategies to derive the RT formula based on the AdS/CFT. The first one is the Rindler method whose physical logic is first proposed in \cite{Casini:2011kv}. Later the authors of \cite{Song:2016gtd,Jiang:2017ecm} find a general way to construct Rindler transformations using the symmetries of the quantum field theory thus generalize the Rindler method to holographic models beyond AdS/CFT. The key point of the Rindler method is to construct a Rindler transformation which is a symmetry of the theory and maps the causal development of a subregion to a thermalized ``Rindler space''. Thus the problem of calculating the entanglement entropy is replaced by the problem of calculating the thermal entropy of the Rindler space. According to holography, the thermal entropy of the Rindler space is given by the thermal entropy of its bulk dual, which is just a hyperbolic  black hole. The horizon of the hyperbolic black hole is exactly what maps to the RT surface under the bulk extended Rindler transformations. An useful by-product of the Rindler method is the picture of boundary and bulk modular flows (see \cite{Jiang:2017ecm}), which play a crucial role in this paper.

The second way is the Lewkowycz-Maldacena (LM) prescription \cite{Lewkowycz:2013nqa} (see \cite{Dong:2016hjy} for its covariant generalization) which extend the replica trick \cite{Calabrese:2004eu} into the bulk, and calculate the entanglement entropy using the on-shell partition function on the replicated bulk geometry. The entanglement entropy is defined as the von Neumann entropy $S_\mathcal{A}=-\text{Tr}\rho_{\mathcal{A}}\log\rho_{\mathcal{A}}$ of the reduced density matrix $\rho_\mathcal{A}$. Consider a quantum field theory on $\mathcal{B}$, the replica trick first calculate the R\'enyi entropy $S^{(n)}_{\mathcal{A}}=\frac{1}{1-n}\log\text{Tr}\rho_{\mathcal{A}}^{n}$ for $n=\mathbb{Z}_{+}$, then analytically continue $n$ away from integers. When $n\to 1$, we get the entanglement entropy $S_{\mathcal{A}}$. To calculate $\rho^{n}_{\mathcal{A}}$, we cut $\mathcal{B}$ open along $\mathcal{A}$, glue n copies of them cyclically, then do path integral on the newly glued manifold $\mathcal{B}_n$. The entanglement entropy is calculated by
\begin{align}
S_{\mathcal{A}}=-n\partial_n \left(\log \mathcal{Z}_n-n\log\mathcal{Z}_1\right)|_{n=1}\,,
\end{align}
where $\mathcal{Z}_n$ is the partition function of the quantum field theory on $\mathcal{B}_n$. Assuming holography and the unbroken replica symmetry in the bulk, the LM prescription manages to construct the bulk dual of $\mathcal{B}_n$, which is a replicated bulk geometry $\mathcal{M}_n$ with its boundary being $\mathcal{B}_n$. Then $\mathcal{Z}_n$ can be calculated by path integral on $\mathcal{M}_n$ on the gravity side. 

In this paper, based on the above two stories, we explore the fine correspondence between the points on the boundary interval $\mathcal{A}$ and the points on the according RT surface $\mathcal{E}_{\mathcal{A}}$ with the guidance of modular flows. Then we relate the fine structure to the entanglement contour, which  characterizes the spatial structure of entanglement entropy.

\section{Boundary and bulk modular flows}
We consider a straight interval
\begin{align}
\mathcal{A}:\{(-\frac{l_u}{2},-\frac{l_v}{2})\to(\frac{l_u}{2},\frac{l_v}{2})\}
\end{align}
at the boundary of the Poincar\'e AdS$_3$
\begin{align}
ds^2=2 r du dv+\frac{dr^2}{4 r^2}\,.
 \end{align}
Here we have set the AdS radius $\ell=1$. We can go back to the usual Poincar\'e coordinates $ds^2=\frac{dx^2+dz^2-dt^2}{dz^2}$ by setting 
\begin{align}
u =\frac{t+x}{2}\,,\quad v = \frac{x-t}{2}\,,\quad r = \frac{2}{z^2}\,.
\end{align}
 The causal development $\mathcal{D}_{\mathcal{A}}$ of $\mathcal{A}$ is given by
\begin{align}
\mathcal{D}_{\mathcal{A}}:~~-\frac{l_u}{2}<u<\frac{l_u}{2}\,,\quad -\frac{l_v}{2}<v<\frac{l_v}{2}\,.
\end{align}
Accordingly the extremal surface $\mathcal{E}_{\mathcal{A}}$ and the corresponding two normal null hypersurfaces $\mathcal{N}_{\pm}$ are given by
\begin{align}
\mathcal{E}_{\mathcal{A}}:&~~~\left\{v= \frac{l_v u}{l_u}\,, ~~r= \frac{2 l_u}{l_u^2 l_v-4 l_v u^2}\,,~~-\frac{l_u}{2}<u<\frac{l_u}{2}\right\}\,,
\\
\mathcal{N}_\pm:&~~~~ r= \frac{2}{(l_u\pm2 u) (l_v\mp2 v)}\,.
\end{align}
The entanglement wedge $\mathcal{W}_{\mathcal{A}}$ \cite{Headrick:2014cta} is the bulk region enclosed by $\mathcal{D}_{\mathcal{A}}$ and $\mathcal{N}_{\pm}$.

Following the strategy in \cite{Jiang:2017ecm}, we can construct a Rindler transformation on the boundary, which is a conformal mapping that maps $\mathcal{D}_{\mathcal{A}}$ to a Rindler space $\tilde{\mathcal{B}}$ with coordinates ($\tilde{u},\tilde{v}$) and infinitely faraway boundary. The strategy requires the translation along the new coordinates to be a linear combination of the global generators of the boundary CFT. Since the global generators are dual to the bulk isometries, we can naturally extend the Rindler transformations into the bulk by replacing the global generators of the CFT with the isometries of AdS$_3$. The bulk extended Rindler transformations map the entanglement wedge $\mathcal{W}_{\mathcal{A}}$ to the exterior of the uncompactified horizon of a Rindler $\widetilde{\text{AdS}}_3$ spacetime with a thermal circle $
(\tilde{u}, \tilde{v})\sim(\tilde{u}+i\tilde{\beta}_{\tilde{u}},\tilde{v}+i\tilde{\beta}_{\tilde{v}})$.

We can write the reduced density matrix as $\rho_{\mathcal{A}}=e^{-H_{\mathcal{A}}}$, where $H_{\mathcal{A}}$ is known as the modular Hamiltonian. The state in the Rindler space $\tilde{B}$ is a thermal state with the thermal density matrix $\rho_{\tilde{B}}=e^{-\beta H_\tau}/Z$, where $Z=\text{tr}\,e^{-\beta H_\tau}$ and $H_\tau$ is the ordinary Hamiltonian in $\tilde{B}$. The modular flow in $\tilde{B}$ is just the ordinary time translation along the thermal circle $k_t=\tilde{\beta}^{i}\partial_{\tilde{x}^i}$. Similarly we can extend $k_t$ into the bulk and get a bulk modular flow $k_t^{bulk}$. With the inverse Rindler transformations, we get the bulk and boundary modular flows,
\begin{align}
k_t=&\left(\frac{2 \pi  u^2}{l_u}-\frac{\pi  l_u}{2}\right)\partial_u+\frac{1}{2} \pi  \left(-\frac{4 v^2}{l_v}+l_v\right)\partial_v
\cr
k_t^{bulk}
=&\left(\frac{2 \pi  u^2}{l_u}-\frac{\pi  l_u}{2}+\frac{\pi }{l_v r}\right)\partial_u+\frac{\pi}{2}   \left(l_v-\frac{2}{l_u r}-\frac{4 v^2}{l_v}\right)\partial_v
\cr
&+4 \pi  r \left(\frac{v}{l_v}-\frac{u}{l_u}\right)\partial_r\,,
\end{align}
which is generated by the modular Hamiltonian $H_{\mathcal{A}}$ in the original Poincar\'e AdS$_3$. We present the details of the Rindler transformations and the derivation of modular flows in Appendix \ref{A}. It is easy to check that $
k_t^{bulk}|_{\mathcal{E}_{\mathcal{A}}}=0$, 
which indicates the extremal surface is the fixed points of the bulk modular flow (or bulk replica symmetry).
\begin{figure}[h] 
   \centering
   \includegraphics[width=0.42\textwidth]{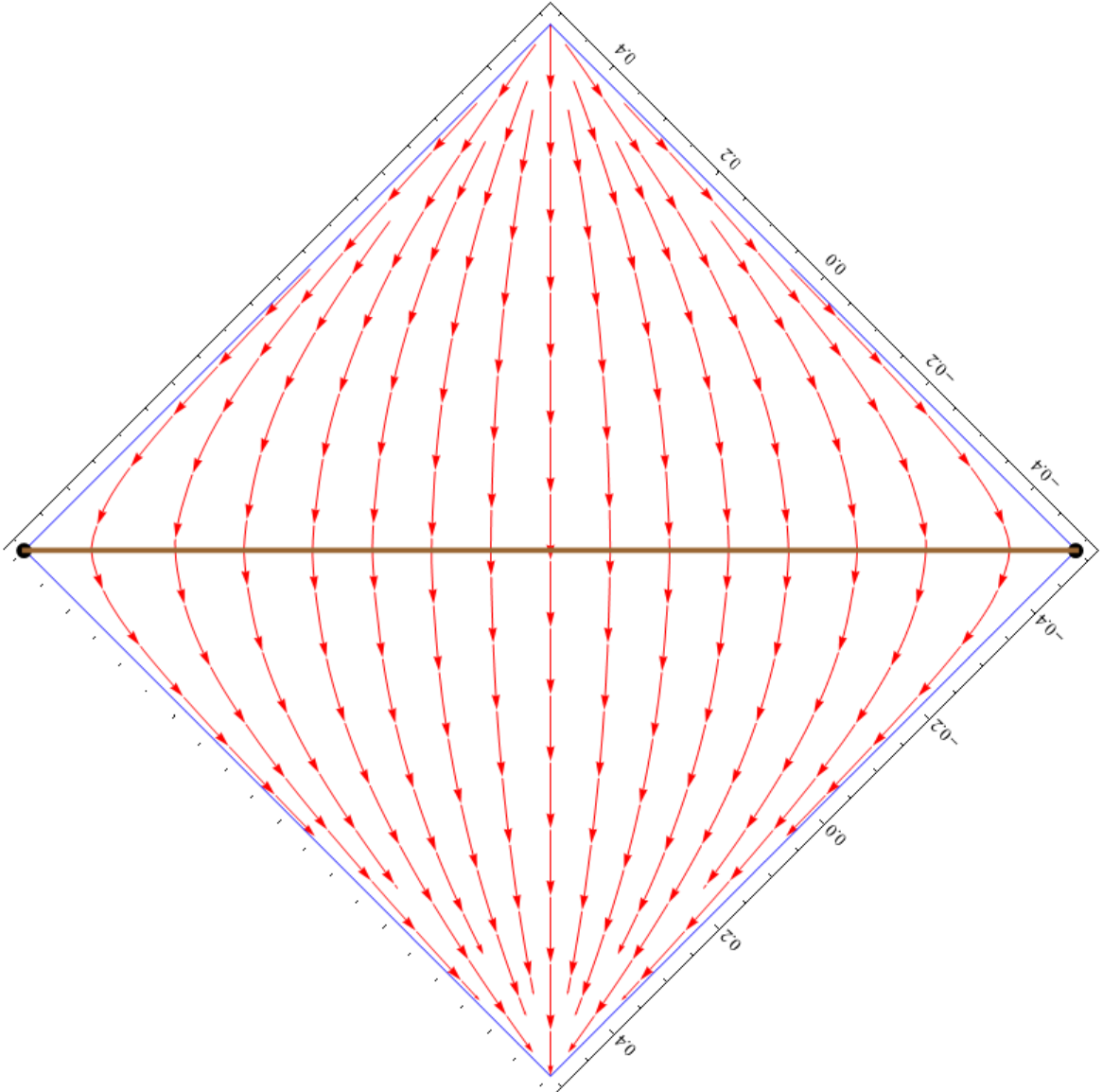} 
   \quad
                \includegraphics[width=0.42\textwidth]{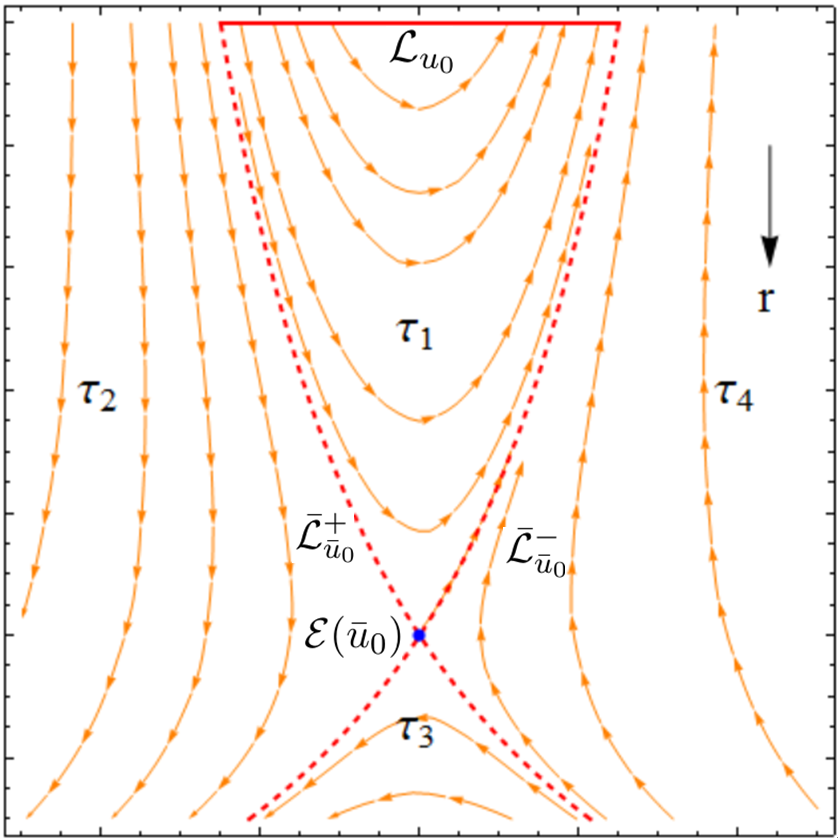} 
 \caption{The left figure shows the modular flow in the causal development $\mathcal{D}_{\mathcal{A}}$ on $\mathcal{B}$. The brown line is the interval $\mathcal{A}$. The right figure shows the modular plane $\mathcal{P}(u_0)$ which is the bulk extension of the boundary modular flow line $\mathcal{L}_{u_0}$. The red and orange arrows depict the direction of the boundary and bulk modular flows respectively.
\label{modularflowcft} }
\end{figure}

 Solving the equations $(\frac{\partial u(s)}{\partial s},\frac{\partial v(s)}{\partial s})=k_t$ and $(\frac{\partial u(s)}{\partial s},\frac{\partial v(s)}{\partial s},\frac{\partial r(s)}{\partial s})=k_t^{bulk}$ respectively, we can get the functions of the modular flow lines both on the boundary and in the bulk. From now on we consider the static case with $l_u=l_v=l/2$. On the boundary, up to a reparametrization, the modular flow lines $\mathcal{L}_{u_0}$ are given by 
\begin{align}\label{boundarymodularflow}
\mathcal{L}_{u_0}:~~\Big\{
\begin{array}{cc}
u(\lambda)&=\frac{l_u}{2} \tanh \left( \tanh ^{-1}(\frac{2 u_0}{l_u})+\log (\lambda )\right)\,,
\\
 v(\lambda)&=\frac{l_u}{2} \tanh \left( \tanh ^{-1}(\frac{2 u_0}{l_u})-\log (\lambda )\right) \,,\\ 
  \end{array} 
\end{align}
where $\lambda$ parametrizes $\mathcal{L}_{u_0}$ and $u_0$ characterize different modular flow lines by being the $u$ coordinate of the point where $\mathcal{L}_{u_0}$ intersect with $\mathcal{A}$ (see the left figure in Fig.\ref{modularflowcft}). In the bulk, one can easily check that the normal null geodesics $\bar{\mathcal{L}}_{\bar{u}_0}^{\pm}$ on $\mathcal{N}_{\pm}$ (which is also studied in \cite{Headrick:2014cta}) are also modular flow lines in the bulk, which are given by (for details see Appendix \ref{B})
\begin{align}\label{bulkmodularflow1}
\bar{\mathcal{L}}_{\bar{u}_0}^{+}:~~\Big\{
\begin{array}{cc}
u(r)&=\frac{l_u}{2}-\frac{l_u-2 \bar{u}_0}{\sqrt{2r} \sqrt{l_u^2-4  \bar{u}_0^2}}\,,
\\
v(r)&=-\frac{l_u}{2}+\frac{2 \bar{u}_0+l_u}{\sqrt{2r} \sqrt{l_u^2-4  \bar{u}_0^2}} \,.\\ 
  \end{array} 
  \\\label{bulkmodularflow2}
  \bar{\mathcal{L}}_{\bar{u}_0}^{-}:~~\Big\{
\begin{array}{cc}
u(r)&=-\frac{l_u}{2}+\frac{2 \bar{u}_0+l_u}{\sqrt{2r} \sqrt{l_u^2-4  \bar{u}_0^2}}\,,
\\
v(r)&=\frac{l_u}{2}-\frac{l_u-2 \bar{u}_0}{\sqrt{2r} \sqrt{l_u^2-4  \bar{u}_0^2}} \,,\\ 
  \end{array} 
\end{align}
where $\bar{u}_0$ characterize different modular flow lines by being the $u$ coordinate of the point where $\bar{\mathcal{L}}_{\bar{u}_0}^{\pm}(r)$ intersect with $\mathcal{E}_{\mathcal{A}}$.

\section{Slicing the entanglement wedge with modular planes}
\begin{figure}[h] 
   \centering
           \includegraphics[width=0.55\textwidth]{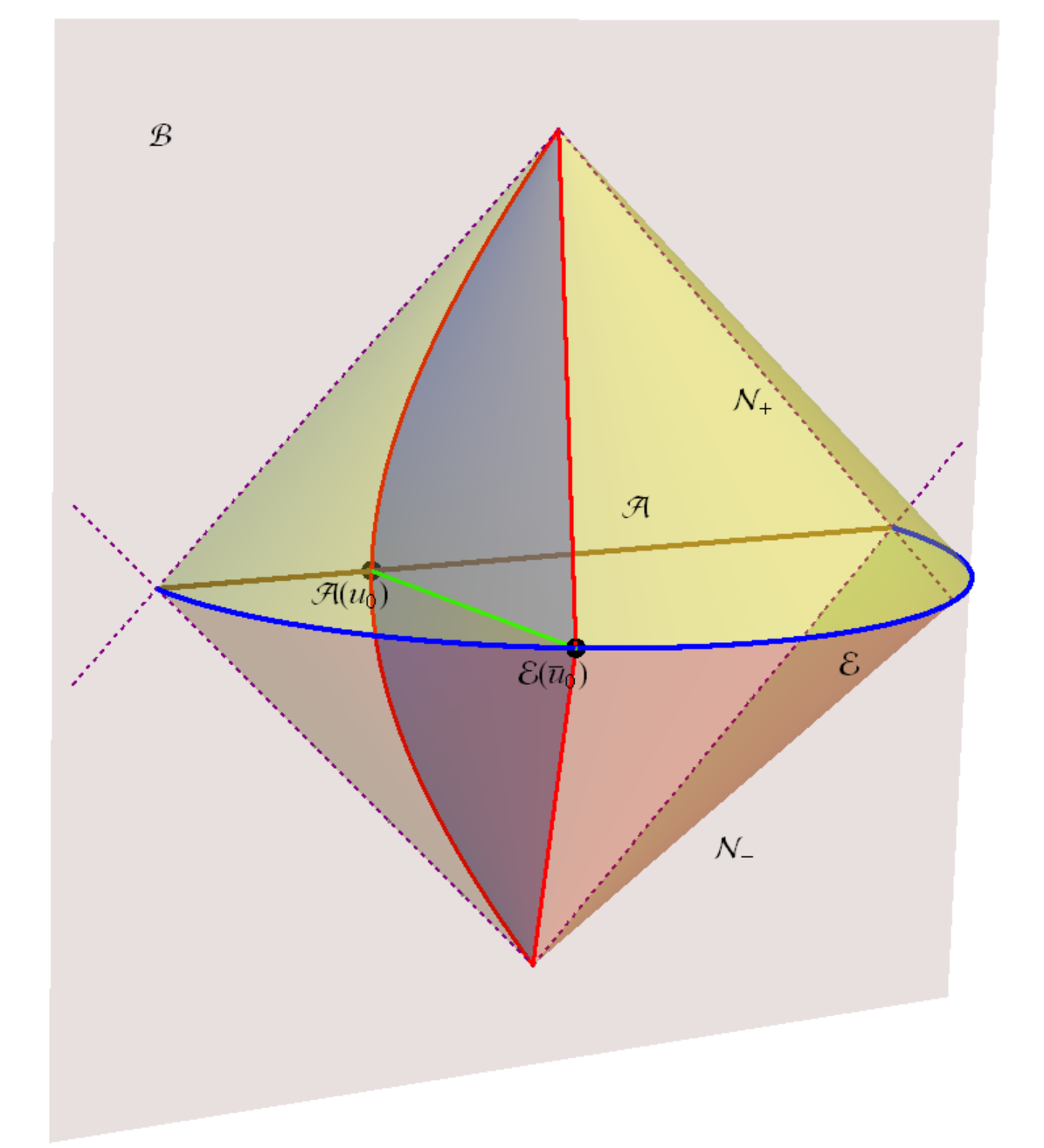}     
 \caption{This figure shows a typical modular plane $\mathcal{P}(u_0)$ in the entanglement wedge. We depict $\mathcal{P}(u_0)$ as the blue surface that intersect with $\mathcal{B}$ and $\mathcal{N}_{\pm}$ on $\mathcal{L}_{u_0},\bar{\mathcal{L}}_{\bar u_0}^{\pm}$, which are depicted as the three red lines, respectively. The green line is $\mathcal{R}_{\mathcal{A}}^{u_0}$ where the modular plane $\mathcal{P}(u_0)$ intersect with $\mathcal{R}_{\mathcal{A}}$.
\label{modularplane} }
\end{figure}
For a given $u=u_0$ with $-l_u/2<u_0<l_u/2$, we define the \textit{modular plane} $\mathcal{P}(u_0)$ as the orbit of the boundary modular flow line $\mathcal{L}_{u_0}$ under the bulk modular flow.  $\mathcal{P}(u_0)$ is a codimension one surface in the bulk (see the right figure in Fig.\ref{modularflowcft}). It is nice to know that for every point $\mathcal{L}_{u_0}(\lambda)$ on $\mathcal{L}_{u_0}$, its orbit under $k_{t}^{bulk}$ will return back to $\mathcal{L}_{u_0}$ on another point, which indicates that $\mathcal{P}(u_0)$ and $\mathcal{L}_{u_0}$ are in one-to-one correspondence. 
With $\mathcal{L}_{u_0}$ and $k_t^{bulk}$ known, the modular plane can be uniquely determined. We define the two points
\begin{align}\label{Au0Eub0}
\mathcal{A}(u_0): (u,v,r)=&(u_0,u_0,\infty)\,,
\cr
 \mathcal{E}(\bar u_0): (u,v,r)=&(\bar{u}_0,\bar{u}_0,\frac{2 }{l_u^2-4  \bar u_0^2})
  \end{align}
as the points where $\mathcal{P}(u_0)$ intersect with $\mathcal{A}$ and $\mathcal{E}_{\mathcal{A}}$ respectively (see Fig.\ref{modularplane}). 
  By definition we have
\begin{align}
\mathcal{P}(u_0)\cap\mathcal{B}=\mathcal{L}_{u_0}\,,\qquad
\mathcal{P}(u_0)\cap\mathcal{N}_{\pm}=\bar{\mathcal{L}}_{\bar{u}_0}^{\pm}\,.
\end{align} 

Define the homology surface $\mathcal{R}_{\mathcal{A}}$ as a codimension one spacelike surface in $\mathcal{W}_{\mathcal{A}}$ which satisfies $\partial\mathcal{R}_{\mathcal{A}}=\mathcal{A}\cup\mathcal{E}_{\mathcal{A}} $. The prescription of \cite{Dong:2016hjy} to construct the corresponding bulk replicated geometry is in the following. Firstly, for each copy of bulk $\mathcal{M}^I$, where $I=1,2,\cdots,n$ denote the $I$th copy of the bulk, we cut them open along $\mathcal{R}^{I}_\mathcal{A}$ to $\mathcal{R}^{I}_\mathcal{A}{}_+$ and $\mathcal{R}^{I}_\mathcal{A}{}_-$. Then we get the replicated geometry by gluing the open cuts cyclically
\begin{align}
\mathcal{R}^{I}_\mathcal{A-}=\mathcal{R}^{(I+1)}_\mathcal{A+}\,,\qquad \mathcal{R}^{n}_\mathcal{A-}=\mathcal{R}^{1}_\mathcal{A+}\,.
\end{align}
Similar to \cite{Dong:2016hjy}, we use bulk Rindler coordinates $\tau_m$ to denote all the bulk regions. We allow $\tau_m$ to be complex and refer to all the bulk regions in question by using
\begin{align}
 \tau_m=\tau+\frac{(m-1)}{2}i\pi\,.
 \end{align} 
 Here $\tau$ parameterizes the modular flow in the bulk with a thermal circle $\tau\sim \tau+2\pi i$. Note that when we translate along a modular flow, only Re$[\tau]$ changes while Im$[\tau]$ is fixed. When we apply the replica trick in the bulk, the orbit of modular flows changes accordingly as well as the distribution of Im$[\tau]$.

\begin{figure}[h] 
   \centering
        \includegraphics[width=0.7\textwidth]{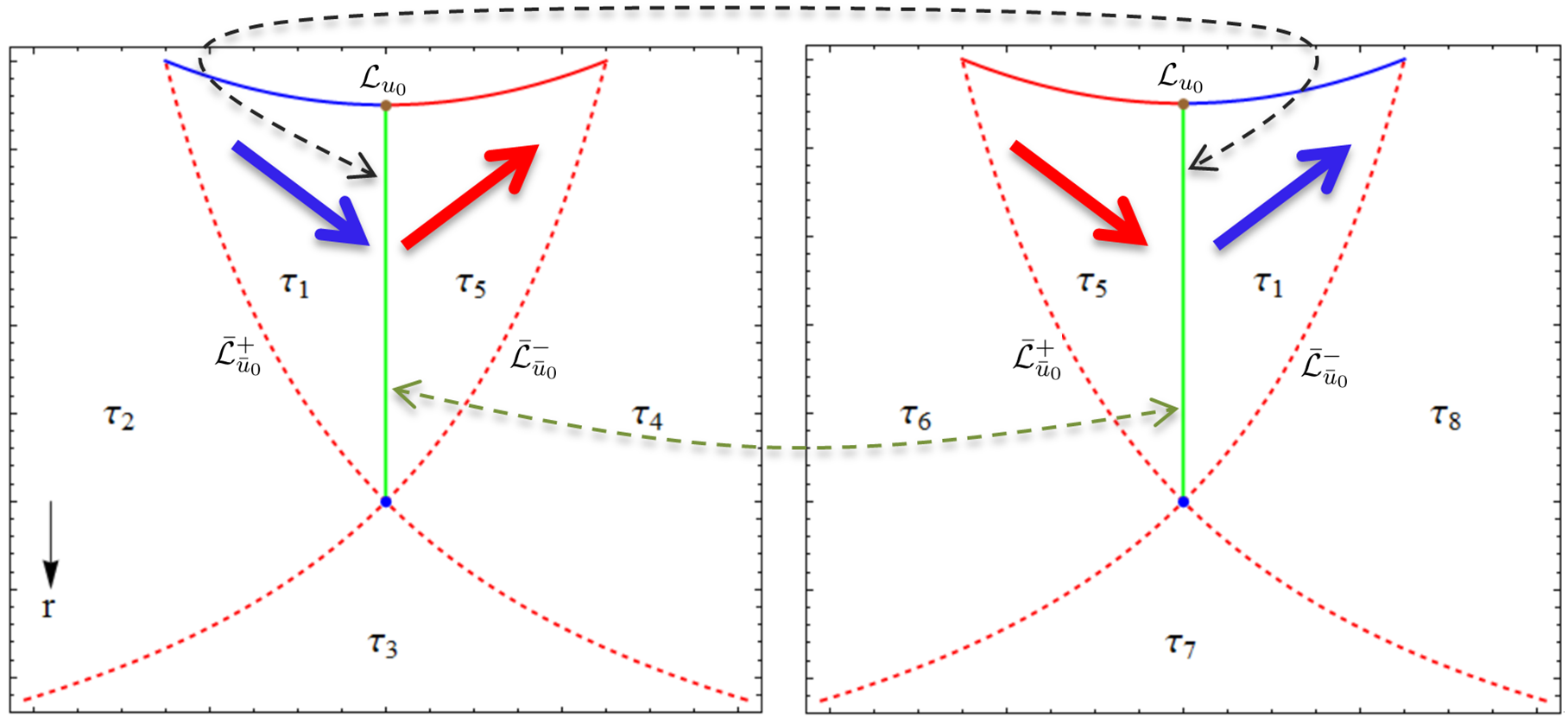}   
 \caption{The replica story on the modular plane $\mathcal{P}(u_0)$ with $n=2$. The left and right figures are the first and second copies of $\mathcal{P}(u_0)$, and the green line is the $\mathcal{R}_{\mathcal{A}}^{u_0}$ which is cut open and glued cyclically. The gluing is depicted by the two dashed arrows. Through $\mathcal{R}_{\mathcal{A}}^{u_0}$, the modular flow $\tau_1$ flows from one subregion of the first copy to a subregion on the second copy (see the blue arrows). The subregion on the second copy should have the same Im$[\tau]$, thus also denoted as $\tau_1$. It is easy to see that on the cyclically glued $\mathcal{P}_2(u_0)$, the thermal circle becomes $\tau_1\to\tau_2\cdots \to \tau_8\to\tau_1$ or in other words $\tau\sim
\tau+4\pi i$.
 \label{bulkreplica} }
\end{figure}

Let us focus on the cyclic gluing of one point $\mathcal{A}(u_0)$ on $\mathcal{A}$.  On the boundary $\mathcal{L}_{u_0}$ passes through $\mathcal{A}(u_0)$ then enter the next copy of $\mathcal{B}$. The natural bulk extension of the cyclic gluing of $\mathcal{A}(u_0)$ should be the cyclic gluing of $\mathcal{R}_{\mathcal{A}}^{u_0}$ on the modular planes, where $\mathcal{R}_{\mathcal{A}}^{u_0}=\mathcal{P}(u_0)\cap \mathcal{R}_{\mathcal{A}}$. In other words, we cut $\mathcal{P}(u_0)$ open along $\mathcal{R}_{\mathcal{A}}^{u_0}$ to $\mathcal{R}_{\mathcal{A}+}^{u_0}$ and $\mathcal{R}_{\mathcal{A}-}^{u_0}$ then impose the following  boundary conditions 
\begin{align}\label{replicaplane}
\psi_I (\mathcal{R}_{\mathcal{A}-}^{u_0})=\psi_{(I+1)}(\mathcal{R}_{\mathcal{A}+}^{u_0})\,,
\quad
\psi_n (\mathcal{R}_{\mathcal{A}-}^{u_0})=\psi_{1}(\mathcal{R}_{\mathcal{A}+}^{u_0})\,,
\end{align}
where $\psi$ denotes all the bulk metric and matter fields. 
Note that the definition of modular plane indicates that all the bulk modular flow lines emanate from $\mathcal{L}_{u_0}$ always lie in $\mathcal{P}(u_0)$. Following the modular flows we can keep track of the value of Im$[\tau]$ everywhere on the cyclically glued modular plane $\mathcal{P}_n(u_0)$ (see Fig.\ref{bulkreplica} for the case of $n=2$). We find that the cyclic gluing of $\mathcal{A}(u_0)$ on the boundary induces a thermal circle $\tau\sim \tau+2\pi n i$ on $\mathcal{P}_{n}(u_0)$. 

In summary, following the modular flow, the cyclic gluing of a point $\mathcal{A}(u_0)$ on the boundary interval effectively induces a replica story on the corresponding modular plane $\mathcal{P}(u_0)$. Following the calculation in \cite{Lewkowycz:2013nqa,Dong:2016hjy}, this turns on nonzero contribution to the entanglement entropy $S_{\mathcal{A}}$ on $\mathcal{E}(\bar u_0)$. The whole bulk replica story can be considered as a slicing of replica stories on all the modular planes. In this sense a point on the boundary interval is related to a unique point on the RT surface. For the specific choice of $\mathcal{A}$ as a straight interval, we find the two points in \eqref{Au0Eub0} are related by (for details see Appendix \ref{B})
\begin{align}\label{ub0u0}
\bar{u}_0=\frac{2 l_u^2 u_0}{4 u_0^2+l_u^2}\,.
\end{align} 

\begin{figure}[h] 
   \centering
        \includegraphics[width=0.42\textwidth]{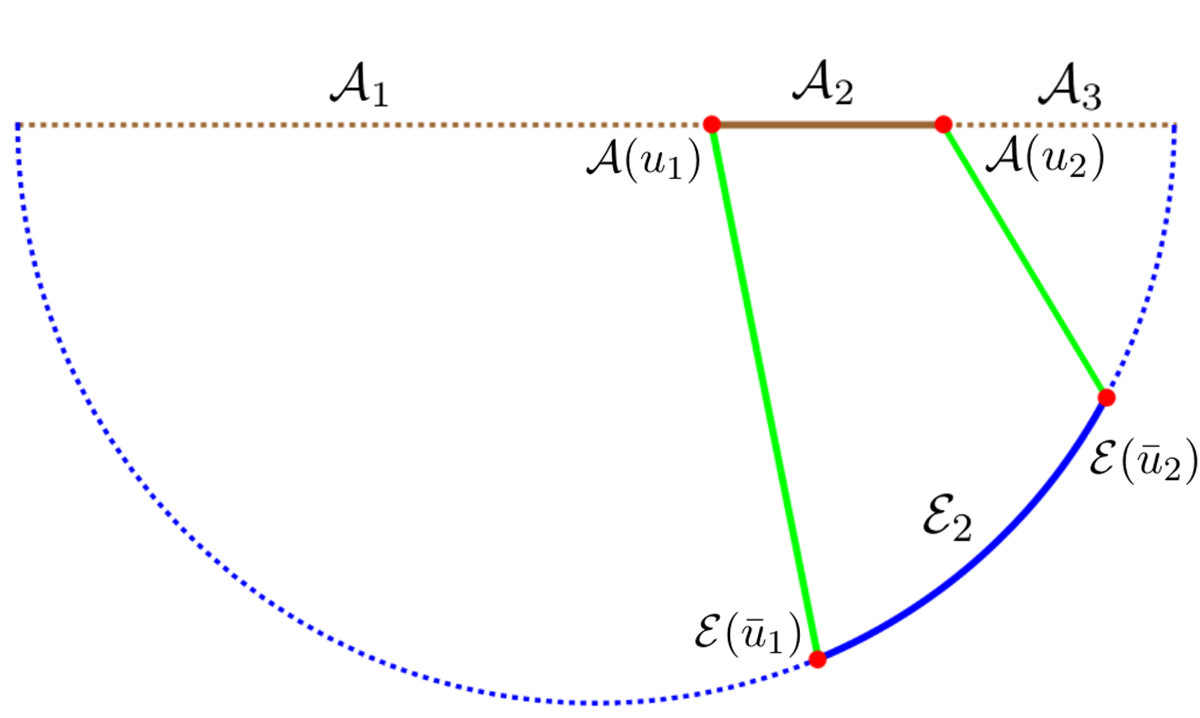}  
        \quad
        \includegraphics[width=0.42\textwidth]{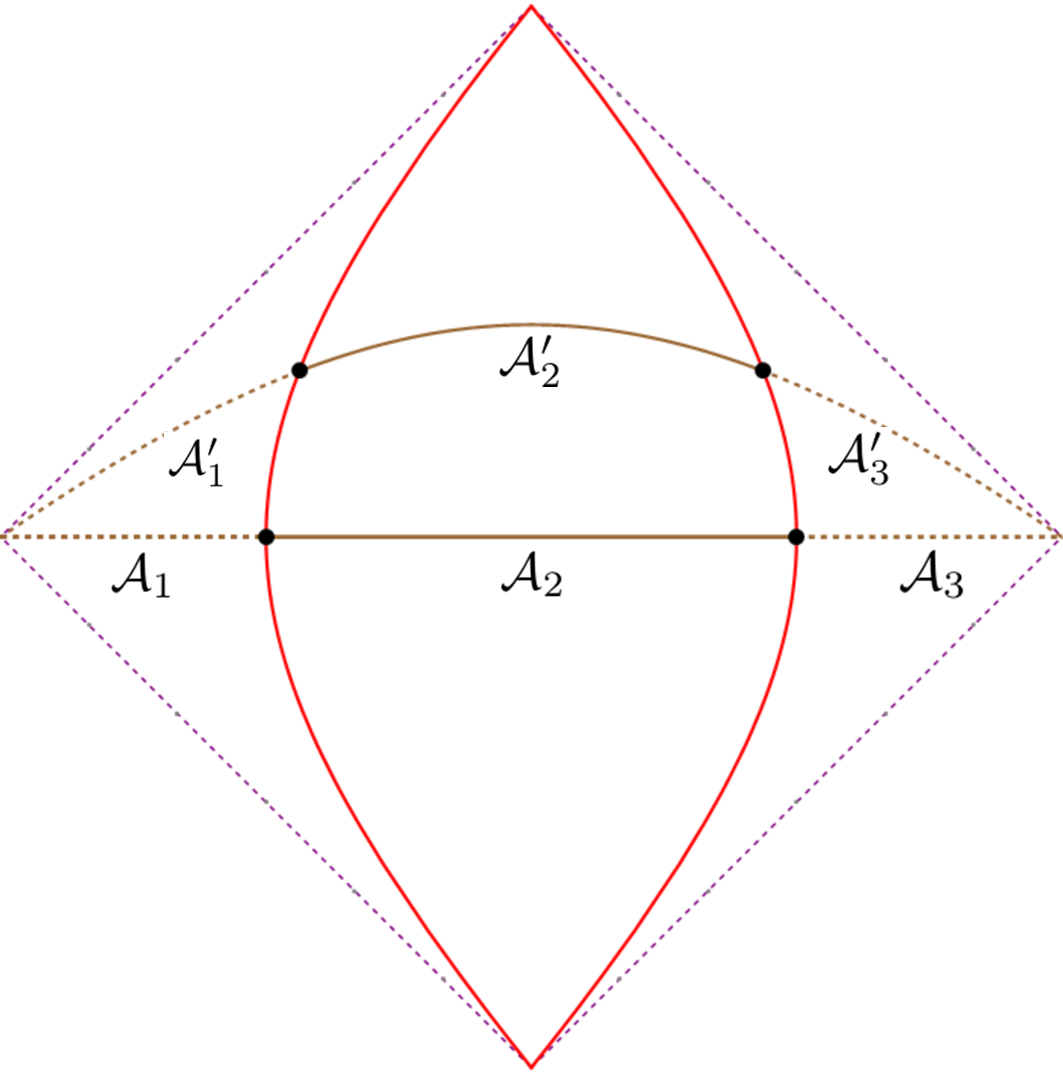}  
 \caption{The left figure shows the subinterval to subinterval correspondence, where the brown line is the boundary interval $\mathcal{A}$ while the blue line is the RT surface $\mathcal{E}_{\mathcal{A}}$. Here $\mathcal{A}$ is divided into three subintervals $\mathcal{A}_1,\mathcal{A}_2$ and $\mathcal{A}_3$ and the two green lines are $\mathcal{R}_{\mathcal{A}}^{u_1}$ and $\mathcal{R}_{\mathcal{A}}^{u_2}$ respectively. The right figure depicts another interval $\mathcal{A}'$ which is homologous to $\mathcal{A}$, and divided into three subintervals $\mathcal{A}'_1,\mathcal{A}'_2$ and $\mathcal{A}'_3$. We require that the endpoints of $\mathcal{A}'_2$ and $\mathcal{A}_2$ are anchored on the same boundary modular flow lines.
\label{aa} }
\end{figure}

We want to address that, according to our prescription the cyclic gluing of an arbitrary point on $\mathcal{L}_{u_0}$ would induce the same replica story on $\mathcal{P}(u_0)$. In other words $\mathcal{E}(\bar{u}_0)$ correspond to all the points on $\mathcal{L}_{u_0}$ in the same sense as $\mathcal{A}(u_0)$.

\section{Entanglement contour from the fine structure}
In the same sense, the cyclic gluing of an arbitrary subinterval $\mathcal{A}_2=\overline{\mathcal{A}(u_1)\mathcal{A}( u_2)}$ on $\mathcal{A}$  turns on the contribution to the entanglement entropy from the subinterval $\mathcal{E}_{2}=\overline{\mathcal{E}(\bar u_1)\mathcal{E}(\bar u_2)}$ of the RT surface (see the left figure of Fig.\ref{aa}). We use $l, l_1, l_2, l_3$ to denote the length of  the intervals $\mathcal{A},\mathcal{A}_1,\mathcal{A}_2$ and $\mathcal{A}_3$ respectively. It is natural to propose that $Length(\mathcal{E}_{2})$ captures the contribution from $\mathcal{A}_2$ to the entanglement entropy $S_\mathcal{A}$. In other words we get the contour function $s_{\mathcal{A}}(x)$ for $S_\mathcal{A}$ which describes the distribution of contribution to entanglement from each point of $\mathcal{A}$ and satisfies
\begin{align}
S_{\mathcal{A}}=\int_{\mathcal{A}}s_{\mathcal{A}}(x)dx\,.
\end{align}

The authors of \cite{Vidal} proposed a set of requirements for the contour functions. Few analysis of the contour functions for bipartite entanglement have been explored in \cite{Botero,Vidal,PhysRevB.92.115129,Coser:2017dtb,Tonni:2017jom}.  However the complete list of requirements that uniquely determines the contour is still not available. Also its fundamental definition is still not established. Our fine structure analysis gives a holographic definition for the contour function.
According to \eqref{ub0u0} we have
\begin{align}
s_{\mathcal{A}}(x)=\frac{1}{4G}\frac{4 l}{l^2-4 x^2}\,,
\end{align}
which is consistent with the results in \cite{Vidal,Coser:2017dtb}. Also we get
\begin{align}\label{newI}
s_{\mathcal{A}}(\mathcal{A}_{2})&=\int_{\mathcal{A}_2}s_{\mathcal{A}}(x)dx=\frac{Length(\mathcal{E}_{2})}{4G}
\\
&=\frac{c}{6}\log\left(\frac{(l_1+l_2)(l_2+l_3)}{l_1 l_3 }\right)\,,\nonumber
\end{align}
where we have used $c=\frac{3\ell}{2G}$. In Appendix \ref{C}, we compare $s_{\mathcal{A}}(\mathcal{A}_{2})$ with the mutual information and show that they are not the same thing.

We consider $\mathcal{A}'$ as an arbitrary spacelike interval homologous to $\mathcal{A}$, and $\mathcal{A}'_{2}$ as the subinterval that ends on the same two modular flow lines $\mathcal{L}_{u_1}$ and $\mathcal{L}_{u_2}$ as $\mathcal{A}_2$ (see the right figure of Fig.\ref{aa}). Since $\mathcal{A}_2$ and $\mathcal{A}'_2$ go through the same modular planes, according to our prescription, they should both correspond to $\mathcal{E}_2$, thus we should have 
\begin{align}\label{A2A2'}
s_{\mathcal{A}}(\mathcal{A}_{2})=s_{\mathcal{A}}(\mathcal{A}_{2}')\,,
\end{align}
which means the entanglement contour is invariant under the boundary modular flow. This requirement should be satisfied in more general cases with locally defined modular Hamiltonians, and is new compared with the requirements in \cite{Vidal}.

It is also interesting to consider the limit $l_1=l_3=\epsilon\to 0$, as expected we find
\begin{align}\label{IS}
S_{\mathcal{A}}=s_{\mathcal{A}}(\mathcal{A}_{2})|_{\mathcal{A}_{2}\to\mathcal{A}}=\frac{c}{3}\log \frac{l}{\epsilon}+\mathcal{O}(\epsilon)\,.
\end{align}
Under this limit the property \eqref{A2A2'} naturally reduce to the causal property $S_\mathcal{A}=S_{\mathcal{A}'}$ of entanglement entropy. Note that the proposal $S_\mathcal{A}$=$\frac{I(\mathcal{A}_{2},\mathcal{A}^{c})}{2}|_{\mathcal{A}_2\to\mathcal{A}}$ \cite{Casini:2008wt} that involves mutual information can not reproduce the right causal property of $S_{\mathcal{A}}$. The points $\mathcal{E}(\bar u_1)$ and $\mathcal{E}(\bar u_2)$, where $\mathcal{E}_2$ is cut off, satisfy $z=\epsilon$, thus relates the boundary and bulk cutoffs in a natural way. This is because the modular planes are defined in a holographic way.

\section{A simple proposal for the contour function}
One interesting observation in our special case is that $s_{\mathcal{A}}(\mathcal{A}_{2})$ can be expressed as a linear combination of the entanglement entropy of single intervals inside $\mathcal{A}$
\begin{align}\label{newIS}
s_{\mathcal{A}}(\mathcal{A}_{2})=\frac{1}{2}\left(S_{\mathcal{A}_1\cup\mathcal{A}_2}+S_{\mathcal{A}_2\cup\mathcal{A}_3}-S_{\mathcal{A}_1}-S_{\mathcal{A}_3}\right)\,.
\end{align}
Here we would like to propose that the above simple combination gives the contour function of entanglement entropy in general 1+1 dimensional theories. We will show that:
\begin{enumerate}
\item $s_{\mathcal{A}}(\mathcal{A}_{2})$ defined by \eqref{newIS} is in general additive by definition, and positive from strong subadditivity \cite{Lieb:1973zz,Lieb:1973cp},
\item as required by \cite{Vidal}, it is invariant under local unitary transformations $U^{\mathcal{A}_2}$ which acts only at the subset $\mathcal{A}_2$, furthermore it is invariant under the modular flow (ie. satisfy \eqref{A2A2'}) in our special case,
\begin{align}\label{SASA'}
S_{\mathcal{A}_1\cup\mathcal{A}_2}+S_{\mathcal{A}_2\cup\mathcal{A}_3}-S_{\mathcal{A}_1}-S_{\mathcal{A}_3}
=S_{\mathcal{A}'_1\cup\mathcal{A}'_2}+S_{\mathcal{A}'_2\cup\mathcal{A}'_3}-S_{\mathcal{A}'_1}-S_{\mathcal{A}'_3}\,,
\end{align}
\item it also reproduces the right contour function for CFT$_2$ with a thermal (spatial) circle, and Warped CFT \cite{Wen:2018mev}.
\end{enumerate} 

Furthermore \cite{Vidal} requires the contour function to satisfy a constraint implementing the consistency with any spatial symmetry of the subsystem, and a bound meaning that $s_{\mathcal{A}}(\mathcal{A}_2)$ must be smaller or equal than the entanglement of any factor space of $\mathcal{H}_{\mathcal{A}}$ which contains the Hilbert space $\mathcal{H}_{\mathcal{A}_2}$ of $\mathcal{A}_2$. We leave these for future discussions.

For quantum systems whose entanglement entropies satisfy the volume law, the proposal \eqref{newIS} gives a flat entanglement contour function. For systems that satisfy area law, \eqref{newIS} also captures the feature that the leading order contribution to the entanglement entropy comes from the boundary.

\subsection{ Additivity and positivity}
We can divide the subinterval $\mathcal{A}_2$ into two parts such that
\begin{align}
\mathcal{A}_{2}=\mathcal{A}^{a}_{2}\cup\mathcal{A}^{b}_{2}\,.
\end{align}
According to \eqref{newIS} we have
\begin{align}
s_{\mathcal{A}}(\mathcal{A}^{a}_{2})&=\frac{1}{2}\left(S_{\mathcal{A}_1\cup\mathcal{A}^{a}_{2}}+S_{\mathcal{A}^{}_{2}\cup\mathcal{A}^{}_{3}}-S_{\mathcal{A}^{}_{1}}-S_{\mathcal{A}^{b}_{2}\cup\mathcal{A}^{}_{3}} \right)\,,
\cr
s_{\mathcal{A}}(\mathcal{A}^{b}_{2})&=\frac{1}{2}\left(S_{\mathcal{A}_1\cup\mathcal{A}^{}_{2}}+S_{\mathcal{A}^{b}_{2}\cup\mathcal{A}^{}_{3}}-S_{\mathcal{A}^{}_{1}\cup\mathcal{A}^{a}_{2}}-S_{\mathcal{A}^{}_{3}} \right)\,,
\end{align}
then we find that
\begin{align}
s_{\mathcal{A}}(\mathcal{A}^{}_{2})=s_{\mathcal{A}}(\mathcal{A}^{a}_{2})+s_{\mathcal{A}}(\mathcal{A}^{b}_{2})\,.
\end{align}
We can continue to do the division such that $\mathcal{A}_{2}$ is divided into all the sites inside $\mathcal{A}_2$, the additivity actually determines a function $s(x)$ on $\mathcal{A}$ that does not depend on the choice of $\mathcal{A}_{2}$ and satisfies
\begin{align}
s_{\mathcal{A}}(\mathcal{A}^{}_{2})=\int_{\mathcal{A}_{2}}s(x)dx\,.
\end{align}
Further more, since 
\begin{align}
S_{\mathcal{A}}=s_{\mathcal{A}}(\mathcal{A}^{}_{2})|_{\mathcal{A}_{2}\to\mathcal{A}}\,,
\end{align}
we have
\begin{align}\label{additivity}
S_{\mathcal{A}}=\int_{\mathcal{A}}s(x)dx\,,
\end{align}
which is a crucial property for the contour function.

\begin{figure}[h] 
   \centering
        \includegraphics[width=0.55\textwidth]{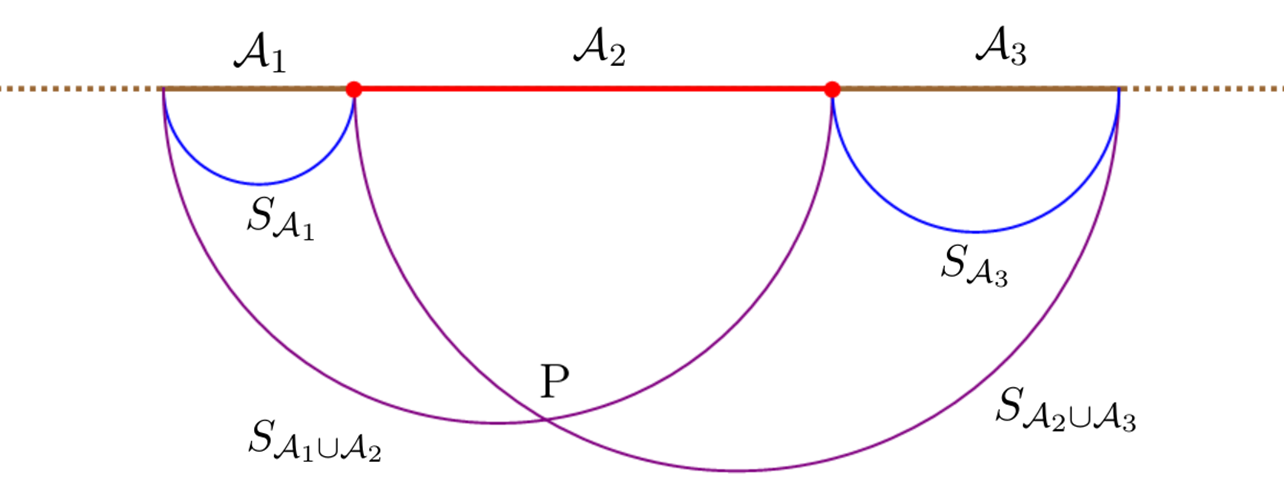}  
 \caption{ The RT surfaces associated to $S_{\mathcal{A}_1\cup\mathcal{A}_2}$ and $S_{\mathcal{A}_2\cup\mathcal{A}_3}$ intersect at the point P. We divide these two RT surfaces by P, then the combination of their left parts is a surface homological to $\mathcal{A}_1$ and its length should be larger than the RT surface associated to $S_{\mathcal{A}_1}$ as $S_{\mathcal{A}_1}$ is minimal. The same logic applies to the combination of the right parts. Then we get \eqref{positivity}.
\label{combb} }
\end{figure}

The positivity of  is directly given by the strong subadditivity. For example, if we consider holographic CFT$_2$ and a static interval $\mathcal{A}$, we see from Fig.\ref{combb} that \cite{Headrick:2007km}
\begin{align}\label{positivity}
S_{\mathcal{A}_1\cup\mathcal{A}_2}+S_{\mathcal{A}_2\cup\mathcal{A}_3}-S_{\mathcal{A}_1}-S_{\mathcal{A}_3}>0\,.
\end{align}
According to \eqref{newIS} and the additivity, the above inequality indicates that the contour function is positive everywhere inside $\mathcal{A}$
\begin{align}\label{positivity1}
s_{\mathcal{A}}(x)>0.
\end{align}

Note that we only used the definition \eqref{newIS} of $s_{\mathcal{A}}(\mathcal{A}_2)$ and the strong subadditivity of entanglement entropy, so the above properties \eqref{additivity} and \eqref{positivity1} of \eqref{newIS} should hold for general cases.

\subsection{Invariance under local unitary transformations and the modular flow}

The causal property of entanglement entropy tells us that $S_{\mathcal{A}_1\cup\mathcal{A}_2}$ and $S_{\mathcal{A}_2\cup\mathcal{A}_3}$ are invariant under local unitary transformations $U^{\mathcal{A}_2}$ that only acts on the subset $\mathcal{A}_2$, then the $s_{\mathcal{A}}(\mathcal{A}_2)$ defined by \eqref{newIS} is also invariant under $U^{\mathcal{A}_2}$ thus satisfies the requirement of \cite{Vidal}. Furthermore, \eqref{newIS} is also invariant under the local unitary transformations $U^{\mathcal{A}_1}$ and $U^{\mathcal{A}_3}$.

Our fine structure analysis indicates that even under modular flow, the contour function should be invariant in the sense of \eqref{A2A2'}. Remarkably we show that the $s_{\mathcal{A}}(\mathcal{A}_2)$ defined by \eqref{newIS} is also invariant under the modular flow in our special case.
According to \eqref{See}, the entanglement entropy for an arbitrary interval is determinant by the coordinate differences $(\Delta u,\Delta v)$  of the two endpoints
\begin{align}\label{See'}
S_{\text{EE}}=\frac{c}{6}\log\frac{\Delta u\Delta v}{\epsilon_u\epsilon_v}\,.
\end{align}
Now we consider an arbitrary spacelike interval $\mathcal{A}'$ which is homologous to $\mathcal{A}$ and intersect with $\mathcal{L}_{u_1}$ and $\mathcal{L}_{u_2}$ at

\begin{align}\label{u1'v1'}
(u'_1,v'_1)=&\left\{\frac{l_u}{2}  \tanh \left(\frac{1}{2} \log \left(\frac{\lambda_1^2 (l_u+2 u_1)}{l_u-2 u_1}\right)\right),\frac{l_u}{2}  \tanh \left(\frac{1}{2} \log \left(\frac{l_u+2 u_1}{\lambda_1^2 l_u-2 \lambda_1^2 u_1}\right)\right)\right\}
\cr
(u'_2,v'_2)=&\left\{\frac{l_u}{2}  \tanh \left(\frac{1}{2} \log \left(\frac{\lambda_2^2 (l_u+2 u_2)}{l_u-2 u_2}\right)\right),\frac{l_u}{2}  \tanh \left(\frac{1}{2} \log \left(\frac{l_u+2 u_2}{\lambda_2^2 l_u-2 \lambda_2^2 u_2}\right)\right)\right\}
\end{align}
Since we have already known the coordinates of the two endpoints of $\mathcal{A}'$, using \eqref{See'} and \eqref{u1'v1'} we can calculate
\begin{align}
&S_{\mathcal{A}'_1\cup\mathcal{A}'_2}+S_{\mathcal{A}'_2\cup\mathcal{A}'_3}-S_{\mathcal{A}'_1}-S_{\mathcal{A}'_3}
\cr
=&\frac{c}{6}\left(\log\frac{(u'_2+l_u/2)(l_u/2-u'_1)}{(u'_1+l_u/1)(l_u/2-u'_2)}+\log\frac{(v'_2+l_v/2)(l_v/2-v'_1)}{(v'_1+l_v/1)(l_v/2-v'_2)}\right)
\cr
=&\frac{c}{3}\log\frac{(l_u-2 u_1) (l_u+2 u_2)}{(l_u+2 u_1) (l_u-2 u_2)}\,,
\cr
=&S_{\mathcal{A}_1\cup\mathcal{A}_2}+S_{\mathcal{A}_2\cup\mathcal{A}_3}-S_{\mathcal{A}_1}-S_{\mathcal{A}_3}
\end{align} 
where we have used $l_u=l_v$ in the third line. Remarkably we find the result is independent of the choice of $\lambda_1$ and $\lambda_2$, thus we derived \eqref{SASA'}.

\subsection{Reproducing the contour function for CFT$_2$ with a thermal (spatial) circle}

Here we consider a CFT$_2$ with the inverse temperature $\beta$. The entanglement entropy for an arbitrary static interval is given by
\begin{align}
S_{\Delta x}=\frac{c}{3}\log\left(\frac{\beta}{\pi \epsilon}\sinh\left(\frac{\pi\Delta x }{\beta}\right)\right)
\end{align}
where $\Delta x$ is the length of the interval. According to \eqref{newIS} we have
\begin{align}\label{13}
&s_{\mathcal{A}}(\mathcal{A}_2)=\frac{c}{3}\log\Big[\sinh\left(\frac{\pi (l_1+l_2)}{\beta}\right)\sinh\left(\frac{\pi(l_2+l_3)}{\beta}\right)
\text{csch}\left(\frac{\pi l_1}{\beta}\right)\text{csch}\left(\frac{\pi l_3}{\beta}\right)\Big]\,,
\end{align}
Define $F(x)=\int s_{\mathcal{A}}(x)dx$, then we should have
\begin{align}\label{14}
s_{\mathcal{A}}(\mathcal{A}_2)=F(x_2)-F(x_1)\,,
\end{align}
where
\begin{align}
 x_1=\frac{1}{2}\left(l_1-l_2-l_3\right)\,,\qquad x_2=\frac{1}{2}\left(l_1+l_2-l_3\right)\,,
 \end{align} 
are the two endpoints of $\mathcal{A}_2$. It is easy to see that \eqref{13} can be written in the form of \eqref{14} with $F(x)$ given by
\begin{align}
F(x)=\frac{c}{3}\log\left( \sinh\left(\frac{\pi (l+2x)}{2\beta}\right)\text{csch}\left(\frac{\pi(l-2x)}{2\beta}\right) \right)\,.
\end{align}
Thus we get the contour function for CFT$_2$ with finite temperature
\begin{align}\label{c17}
s_{\mathcal{A}}(x)=\frac{c}{3}\frac{\pi\sinh\frac{l\pi}{\beta}}{\beta\sinh\left(\frac{\pi(l-2x)}{2\beta}\right)\sinh\left(\frac{\pi(l+2x)}{2\beta}\right)}\,.
\end{align}
Similarly we can get the contour function for the zero temperature CFT$_2$ with a spatial circle $x\sim x+L$,
\begin{align}\label{c18}
s_{\mathcal{A}}(x)=\frac{c}{3}\frac{\pi\sin\frac{l\pi}{L}}{L \sin\left(\frac{\pi(l-2x)}{2L}\right)\sin\left(\frac{\pi(l+2x)}{2L}\right)}\,.
\end{align}
The above results \eqref{c17} and \eqref{c18} are consistent with the results proposed in \cite{Coser:2017dtb}, which are also inspired by the locally defined modular Hamiltonian \cite{Bisognano:1975ih,Bisognano:1976za,Casini:2011kv,Cardy:2016fqc}.

\section{Discussion}
The modular Hamiltonian absolutely contains more information than the entanglement entropy. This work shows that a fine correspondence between quantum entanglement and space-time geometry can be extracted from the RT formula by the bulk and boundary modular flows. This fine correspondence is indeed the holographic picture of the entanglement contour, which probes the locality of entanglement and gives more information (see discussions in \cite{Vidal}) than the total entanglement entorpy. Our prescription, although it relies on the construction of Rindler transformations, should work for general cases with a locally defined modular Hamiltonian, for example, the covariant case by setting $l_u\neq l_v$, the global AdS$_3$, AdS$_3$ with a black hole, AdS space in higher dimensions (for spherical subregions Rindler transformations has been constructed in \cite{Casini:2011kv}), and even the cases \cite{Song:2016gtd,Jiang:2017ecm} beyond AdS/CFT (see \cite{Wen:2018mev} for another explicit case in the context of warped AdS$_3$/warped CFT correspondence \cite{Detournay:2012pc}). For cases with nonlocal modular Hamiltonian, one may find clues from the more general discussions on bulk and boundary modular flows in \cite{Faulkner:2013ana,Jafferis:2015del,Faulkner:2017vdd} to define the modular planes in a more abstract way. 

On the other hand we give a simple proposal \eqref{newIS} for the entanglement contour function for general cases, which only involves entanglement entropies of single subintervals inside $\mathcal{A}$ and does not depend on the construction of the Rindler transformations. It also passes several nontrivial tests. The fine correspondence in holographic entanglement together with the proposal \eqref{newIS} allow us to interpret the length of bulk intervals in terms of entanglement entropies of the dual field theory (see \cite{Abt:2018ywl} for a recent application along this line). On the other way around, we can get the information of the modular flow from the entanglement entropy based on this proposal.

The fine structure also gives a good explanation \cite{Wen:2018mev} for the appearance of null geodesics in the new geometric pictures \cite{Song:2016gtd,Jiang:2017ecm} of entanglement entropy in non-AdS holographies.  The RT formula has a very deep impact on our understanding of holography itself and the origin of spacetime geometry \cite{Swingle:2009bg,VanRaamsdonk:2009ar,VanRaamsdonk:2010pw,Maldacena:2013xja}. Its fine description{\footnote{Another attempt to obtain a fine description of the RT formula is given by \cite{Freedman:2016zud} using a tool named ``bit threads'' which represent entanglement between points on the boundary.}} may help us better understand these grand questions.

\section*{ACKNOWLEDGMENTS}
We thank Ping Gao, Temple He, Hongliang Jiang, Aitor Lewkowycz and Mukund Rangamani for helpful discussions. Especially we would like to thank Matthew Headrick and Wei Song for insightful discussions. We also thank Shiu-Yuen Cheng, Wei Song, Shing-Tung Yau and Pin Yu for support. This work is supported by NSFC Grant No.11805109.

\appendix

\section{Rindler method and modular flows in AdS$_3$/CFT$_2$}\label{A}

The general strategy to construct Rindler transformations and their bulk extensions by using the symmetries of a QFT and holographic dictionary, is summarised in Sec. 2 of \cite{Jiang:2017ecm}. In the case of AdS$_3$/CFT$_2$, the Rindler transformations are constructed by imposing the following requirements
\begin{itemize}
\item The Rindler transformation $\tilde{x}=f(x)$ should be a symmetry transformation, which, in this case, is a conformal mapping
\begin{align}\label{Rindler1}
\tilde{u}=f(u)\,,\qquad \tilde{v}=g(v)\,,
\end{align}
with $f$ and $g$ being arbitrary functions.

\item  The vectors $\p_{\tilde{x}^i}$ should be a linear combination of the global generators $h_i$ in the original CFT. In other words
\begin{align}\label{Rindler2}
\partial_{\tilde{x}^i}=\sum_j b_{ij} h_j\,,
\end{align}
where $b_{ij}$ are arbitrary constants.

\item  The bulk extension of the Rindler transformation is obtained by replacing the global generators $h_j$ in \eqref{Rindler2} with their bulk duals, which are just the isometries of the Poincar\'e AdS$_3$. Furthermore we require the metric of the Rindler $\widetilde{\text{AdS}}_3$ to satisfy the same boundary conditions.
\end{itemize}

The global generators $h_i$ of the boundary CFT$_2$ are $L_{0,\pm}$ and $\bar L_{0,\pm}$, whose bulk dual are the isometries of the  Poincar\'e AdS$_3$
\begin{align}\label{sixkv}
J_-=  \partial _u, \quad J_0=  u \p_{u}-r  \p_{r }, \quad J_+=  u^2\p_{u}-\frac{1}{2 r }\p_{v}-2 r  u \p_{r },
\cr
\bar{J}_-=  \partial _v, \quad \bar{J}_0=  v\p_{v}-r \p_{r }, \quad \bar{J}_+=  v^2\p_{v}-\frac{1}{2r }\p_{u}-2 r  v\p_{r }.
\end{align}
The normalization are chosen to satisfy the standard $SL(2,R)\times SL(2,R)$ algebra
\begin{align}
&[J_-,J_+]=2J_0,\quad [J_0,J_\pm]=\pm J_{\pm},
\cr
&[\bar{J}_-,\bar{J}_+]=2\bar{J}_0,\quad [\bar{J}_0,\bar{J}_\pm]=\pm \bar{J}_{\pm}\,.
\end{align}

Now we try to construct a Rindler coordinate transformation that satisfies the above requirements, to obtain a new coordinate system. Define  
\begin{align}
&\partial_{\tilde{u}}=a_0 J_0+a_+ J_++a_- J_-\,,
\cr
&\partial_{\tilde{v}}=\bar{a}_0 \bar{J}_0+\bar{a}_+ \bar{J}_++\bar{a}_- \bar{J}_-\,,
\end{align}
where
 $a_0,a_+,a_-,\bar{a}_0,\bar{a}_+,\bar{a}_-$ are arbitrary constants which controls the size, position of $\mathcal{D}_{\mathcal{A}}$ on $\mathcal{B}$ and the two parameters $\tilde{\beta}_{\tilde{u}}, \tilde{\beta}_{\tilde{v}}$ that characterize the thermal circle in $\tilde{\mathcal{B}}$. Note that the shape of $\mathcal{D}_{\mathcal{A}}$ is determined by the symmetries of the CFT thus can not be adjusted. Only the two parameters that characterize the size of $\mathcal{D}_{\mathcal{A}}$ can affect the entanglement entropy. Furthermore by requiring the metric of the Rindler $\widetilde{\text{AdS}_{3}}$ to have the formula of \eqref{5.2} , or equivalently $\tilde{r}\equiv g_{\tilde{u}\tilde{v}}$, determines the other coordinate $\tilde{r}$.

For simplicity we can settle down the position of $\mathcal{D}_{\mathcal{A}}$ and the thermal circle of the Rindler $\widetilde{\text{AdS}_{3}}$. This leaves only two parameters $l_u$ and $l_v$, which characterize the size of $\mathcal{D}_{\mathcal{A}}$. By choosing 
\begin{align}
a_0=0\,,\quad a_+=-\frac{2}{l_u}\,,\quad a_-=\frac{l_u}{2}\,,
\cr
\bar a_0=0\,,\quad \bar a_+=-\frac{2}{l_v}\,,\quad \bar a_-=\frac{l_v}{2}\,,
\end{align}
we find the bulk Rindler transformations from Poincar\'e AdS$_3$
\begin{align}\label{5.1}
ds^2=2 r du dv+\frac{dr^2}{4r^2}\,,
\end{align}
 to a Rindler $\widetilde{\text{AdS}_3}$
 \begin{align}\label{5.2}
 ds^2= d\tilde{u}^2+2 \tilde{r}d\tilde{u}d\tilde{v}+d\tilde{v}^2+\frac{d\tilde{r}^2}{4(\tilde{r}^2-1)}\,,
 \end{align}
 are given by
\begin{align}\label{rindlerads}
\tilde{u}&=\frac{1}{4} \log \left(\frac{4 (r v (l_u+2 u)+1)^2-l_v^2 r^2 (l_u+2 u)^2}{4 (r v (l_u-2 u)-1)^2-l_v^2 r^2 (l_u-2 u)^2}\right),
\\
\tilde{v}&=\frac{1}{4} \log \left(\frac{l_u^2 r^2 (l_v+2 v)^2-4 (l_v r u+2 r u v+1)^2}{l_u^2 r^2 (l_v-2 v)^2-4 (-l_v r u+2 r u v+1)^2}\right),
\\\label{trr}
\tilde{r}&=\frac{r^2 \left(l_u^2 \left(l_v^2-4 v^2\right)-4 l_v^2 u^2\right)+4 (2 r u v+1)^2}{4 l_u l_v r}.
\end{align}
Here we have set down the parameters $\tilde{\beta}_{\tilde{u}}=-\tilde{\beta}_{\tilde{v}}=-\pi$ and the position of $\mathcal{D}_{\mathcal{A}}$, which do not affect the entanglement entropy. Asymptotically, we have
\begin{align}
\tilde{u}=\text{arctanh} \frac{2 u}{l_u}\,,\qquad \tilde{v}=\text{arctanh} \frac{2 v}{l_v}\,,
\end{align}
which is as expected a conformal mapping. The $(\tilde{u},\tilde{v})$ coordinates covers a diamond shape subregion $\mathcal{D}_{\mathcal{A}}$ on the original $(u,v)$ coordinates
\begin{align}\label{DA}
\mathcal{D}_{\mathcal{A}}:~~-\frac{l_u}{2}<u<\frac{l_u}{2}\,,\quad -\frac{l_v}{2}<v<\frac{l_v}{2}\,.
\end{align}
which is the causal development of the interval  $\mathcal{A}$
\begin{align}\label{tu0tv1}
\mathcal{A}:~~\{(-\frac{l_u}{2},-\frac{l_v}{2})\to(\frac{l_u}{2},\frac{l_v}{2})\}\,,
\end{align}
on the boundary CFT. The causal development \eqref{DA} constructed by only using Rindler transformations is consistent with the causal development $\mathcal{D}_{\mathcal{A}}$  defined with null lines associated to $\partial\mathcal{A}$ on $\mathcal{B}$.

According to \eqref{trr}, the horizon of the Rindler $\widetilde{\text{AdS}_3}$ at $\tilde{r}=1$ maps to two null hypersurfaces $\mathcal{N}_\pm$ in the original space
\begin{align}\label{Nads}
\mathcal{N}_+:~~ r= \frac{2}{(l_u+2 u) (l_v-2 v)}\,,
\quad
 \mathcal{N}_-:~~ r= \frac{2}{(l_u-2 u) (l_v+2 v)}\,.
\end{align}
We see that $\mathcal{N}_\pm$ intersect at 
\begin{align}\label{gammaads}
\mathcal{E}_{\mathcal{A}}:&~\left\{v= \frac{l_v u}{l_u}\,, ~~r= \frac{2 l_u}{l_u^2 l_v-4 l_v u^2}\,,~~-\frac{l_u}{2}<u<\frac{l_u}{2}\right\}.
\end{align}
which anchors on the boundary endpoints $\partial\mathcal{A}_\pm=\left(\pm \frac{l_u}{2},\pm \frac{l_v}{2}\right)$. It is easy to check that the $\mathcal{E}$ is just the extremal surface. The entanglement entropy is given by
\begin{align}\label{See}
S_{\mathcal{A}}=\frac{1}{4G}\log\frac{l_ul_v}{\epsilon_u \epsilon_v}\,,
\end{align}
where $\epsilon_u$ and $\epsilon_v$ are the cutoffs along the $u$ and $v$ directions.

\begin{figure}[h] 
   \centering
        \includegraphics[width=0.42\textwidth]{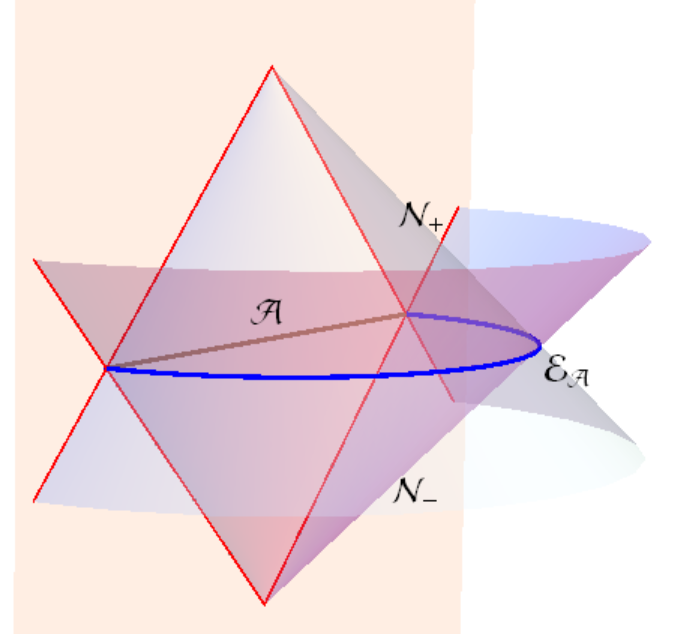}  
        \quad
               \includegraphics[width=0.42\textwidth]{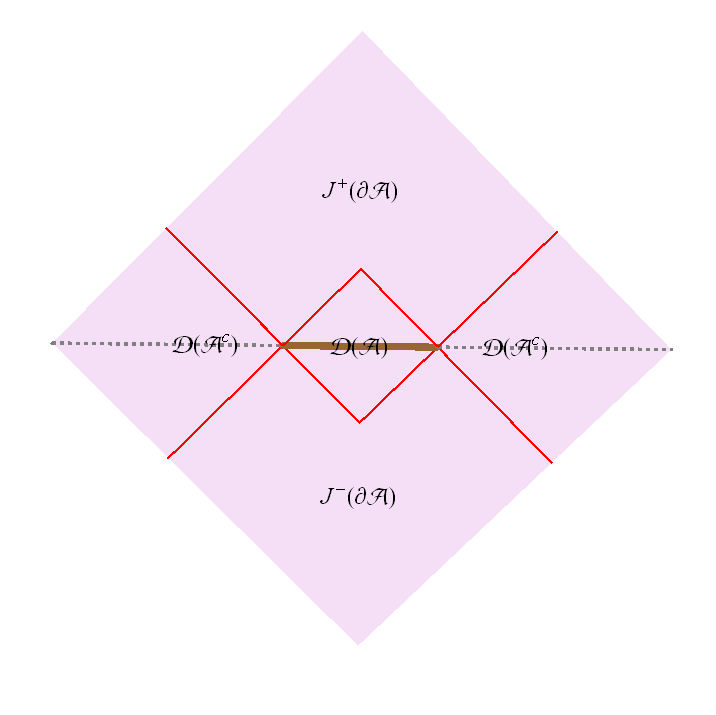}  
 \caption{ The left figure shows the causal decomposition for AdS$_3$ associated with a RT surface. The right figure shows the causal structure associated with an interval $\mathcal{A}$ of CFT$_2$.
\label{adsdecoposition} }
\end{figure}

On the other hand, one can also check that the normal null hypersurfaces emanating from $\mathcal{E}_{\mathcal{A}}$ \eqref{gammaads} are just $\mathcal{N}_\pm$ \eqref{Nads}. This means the normal null hypersurfaces $\mathcal{N}_{\pm}$ of $\mathcal{E}_{\mathcal{A}}$ play the role of the horizon in Rindler $\widetilde{\text{AdS}_3}$. The above picture is just the light-sheet \cite{Bousso:2002ju} construction of the HRT surface first proposed in \cite{Hubeny:2007xt}. The entanglement wedge $\mathcal{W}_\mathcal{A}$ is the bulk region enclosed by $\mathcal{N}_\pm$ and $\mathcal{B}$. The Rindler transformation \eqref{rindlerads} maps this $\mathcal{W}_\mathcal{A}$ to the exterior of the horizon in Rindler  $\widetilde{\text{AdS}_3}$. The bulk causal decomposition associate with $\mathcal{E}_{\mathcal{A}}$ is given by the left figure in Fig.\ref{adsdecoposition}. It is easy to see that the followed boundary decomposition is consistent with the causal structure for a CFT$_2$, which is given by the right figure of Fig.\ref{adsdecoposition}. 

The generator of the normal Hamiltonian in Rindler space or Rindler bulk \eqref{5.2}, which maps to the modular Hamiltonian in the original space, is the generator along the thermal circle $k_t\equiv {\tilde \beta}^i \p_{\tilde{x}^i}$. In order to map it to the original space, we need to solve the following differential equations
\begin{align}\label{diffeq}
\partial_u&=(\partial_u \tilde{u})\partial_{\tilde{u}}+(\partial_u \tilde{v})\partial_{\tilde{v}}+(\partial_u \tilde{r})\partial_{\tilde{r}}\,,
\cr
\partial_v&=(\partial_v \tilde{u})\partial_{\tilde{u}}+(\partial_v \tilde{v})\partial_{\tilde{v}}+(\partial_v \tilde{r})\partial_{\tilde{r}}\,,
\cr
\partial_r&=(\partial_r \tilde{u})\partial_{\tilde{u}}+(\partial_r \tilde{v})\partial_{\tilde{v}}+(\partial_r \tilde{r})\partial_{\tilde{r}}\,.
\end{align}
Then we get $\partial_{\tilde{u}},\partial_{\tilde{v}},\partial_{\tilde{r}}$, and furthermore $k_t$, in terms of $\partial_u,\partial_v,\partial_r$.

We plug the bulk Rindler transformations \eqref{rindlerads} into the differential equations \eqref{diffeq} then solve them. In Rindler  $\widetilde{\text{AdS}_3}$ the generator of the Hamiltonian is just the translation along the thermal circle. Mapping it to the original space,  we get the modular flow in the bulk 
\begin{align}
k_t^{bulk}=&\tilde{\beta}_{\tilde{u}}\partial_{\tilde{u}}+\tilde{\beta}_{\tilde{v}}\partial_{\tilde{v}}
\cr
=&\pi\left(\partial_{\tilde{v}}-\partial_{\tilde{u}}\right)
\cr
=&\left(\frac{2 \pi  u^2}{l_u}-\frac{\pi  l_u}{2}+\frac{\pi }{l_v r}\right)\partial_u+4 \pi  r \left(\frac{v}{l_v}-\frac{u}{l_u}\right)\partial_r
\cr
&+\frac{1}{2} \pi  \left(-\frac{2}{l_u r}-\frac{4 v^2}{l_v}+l_v\right)\partial_v
&\,.
\end{align}
The modular flow on the boundary is given by
\begin{align}
k_t=\left(\frac{2 \pi  u^2}{l_u}-\frac{\pi  l_u}{2}\right)\partial_u+\frac{1}{2} \pi  \left(-\frac{4 v^2}{l_v}+l_v\right)\partial_v\,.
\end{align}
One can easily check that
\begin{align}
k_t|_{\partial\mathcal{A}_\pm}=0\,,\qquad k_t^{bulk}|_{\mathcal{E}_{\mathcal{A}}}=0\,.
\end{align}
The above equations mean that $\mathcal{E}_{\mathcal{A}}$ is the fixed points of $k_{t}^{bulk}$ (or the bulk extended replica symmetry), and  $\mathcal{E}_{\mathcal{A}}$ should go through the endpoints $\partial{\mathcal{A}}_{\pm}$ of the boundary interval.

\section{Details to derive the equations (11) (12) and (18) }\label{B}

\subsection{Normal null geodesics on $\mathcal{N}_{\pm}$ as bulk modular flow lines}

One can check that the following bulk flow lines
\begin{align}
\left\{u(s),v(s),r(s)\right\}=\left\{\frac{l_u \left(b \left(e^{2 \pi  s }-b\right)+1\right)}{2 \left(-b^2+e^{2 \pi  s }+1\right)},\frac{l_u \left(b \left(b+e^{2 \pi  s }\right)-1\right)}{2 \left(-b^2+e^{2 \pi  s }+1\right)},-\frac{2 e^{-4 \pi  s } \left(-b^2+e^{2 \pi  s }+1\right)^2}{\left(b^2-1\right) l_u^2}\right\}
\end{align}
satisfy the equation $(\frac{\partial u(s)}{\partial s},\frac{\partial v(s)}{\partial s},\frac{\partial r(s)}{\partial s})=k_t^{bulk}$, where $b$ is an integration constant. These are the bulk modular flow lines on $\mathcal{N}_{+}$, one can also check that they are the null geodesics normal to $\mathcal{E}_{\mathcal{A}}$, which are also studied in \cite{Headrick:2014cta}. 

We define
\begin{align}
b= \frac{2 \bar{u}_0}{l_u},\qquad e^{-2 \pi  s }= \frac{2 l_u^2-8 \bar{u}_0^2+\sqrt{2} \sqrt{ r \left(l_u^2-4 \bar{u}_0^2\right)^3}}{l_u^2 \left(l_u^2 r-4 r \bar{u}_0^2-2\right)}\,,
\end{align}
then we reparametrize the null bulk modular flow lines with $r$ and $\bar{u}_0$
\begin{align}
\bar{\mathcal{L}}_{\bar{u}_0}^{+}:~~\Big\{
\begin{array}{cc}
u(r)&=\frac{l_u}{2}-\frac{l_u-2 \bar{u}_0}{\sqrt{2r} \sqrt{l_u^2-4  \bar{u}_0^2}}\,,
\\
v(r)&=-\frac{l_u}{2}+\frac{2 \bar{u}_0+l_u}{\sqrt{2r} \sqrt{l_u^2-4  \bar{u}_0^2}} \,.\\ 
  \end{array} 
\end{align}
To get the null bulk modular flow lines on $\mathcal{N}_{-}$, we can just map $(u,v)$ to $(-u,-v)$, thus we derived  \eqref{bulkmodularflow1} and \eqref{bulkmodularflow2}.

\subsection{Deriving equation \eqref{ub0u0}}
On the real boundary $r=\infty$, all the bulk modular flow lines \eqref{bulkmodularflow1} intersect with the boundary modular flow lines \eqref{boundarymodularflow} at the tips $(u,v)=(\pm\frac{l_u}{2},\mp\frac{l_v}{2})$ of the casual development $\mathcal{D}_{\mathcal{A}}$. Our construction of modular planes indicates that for each boundary modular flow line $\mathcal{L}_{u_0}$, there are two bulk modular flow lines $\bar{\mathcal{L}}^{\pm}_{\bar u_0}$ on $\mathcal{N}_{\pm}$ that lie in the same modular plane $\mathcal{P}(u_0)$. To get the relation \eqref{ub0u0}, we need to push the boundary with modular flows \eqref{boundarymodularflow} to a finite $r=r_{I}$, then for each $\mathcal{L}_{u_0}$ there will be a $\bar{\mathcal{L}}_{\bar{u}_0}^{+}$ that intersect with $\mathcal{L}_{u_0}$ at some finite $\lambda$. More explicitly we solve the equation
\begin{align}
\frac{l_u}{2}  \tanh \left(\tanh ^{-1}\left(\frac{2 u_0}{l_u}\right)+\log (\lambda )\right)&=\frac{l_u}{2}-\frac{l_u-2 \bar{u}_0}{\sqrt{2r} \sqrt{l_u^2-4  \bar{u}_0^2}}\,,
\cr
\frac{ l_u}{2} \tanh \left(\tanh ^{-1}\left(\frac{2 u_0}{l_u}\right)-\log (\lambda )\right)&=-\frac{l_u}{2}+\frac{l_u+2 \bar{u}_0}{\sqrt{2r} \sqrt{l_u^2-4  \bar{u}_0^2}} \,,
\end{align}
and get the intersecting point at
\begin{align}
\bar{u}_0=& \frac{2 \lambda^2  l_u^2 u_0}{(\lambda^2 +1) l_u^2+4 (\lambda^2 -1) u_0^2}\,,
\cr
r_I=& \frac{\lambda ^2 \left(\cosh (2 \log (\lambda ))+\cosh \left(2 \tanh ^{-1}\left(\frac{2 u_0}{l_u}\right)\right)\right)}{l_u^2}\,.
\end{align}
Then we take the limit $\lambda\to\infty$ and get
\begin{align}
\bar u_0&=\frac{2 l_u^2 u_0}{l_u^2+4 u_0^2}+\mathcal{O}\left(\frac{1}{\lambda^2}\right)\,,
\\
r_I&=\frac{\lambda ^4}{2 l_u^2}+\frac{\lambda ^2 \cosh \left(2 \tanh ^{-1}\left(\frac{2 u_0}{l_u}\right)\right)}{l_u^2}+\mathcal{O}(1)\,,
\end{align}
which give the relation \eqref{ub0u0}.

\section{Entanglement contour compared with mutual information}\label{C}
We define $\mathcal{I}(\mathcal{A}_2,\mathcal{A}^{c})=2s_{\mathcal{A}}(\mathcal{A}_{2})$  and compare it with the mutual information
\begin{align}\label{mutual}
I(\mathcal{A}_2,\mathcal{A}^c)&=S_{\mathcal{A}_2}+S_{\mathcal{A}^c}-S_{\mathcal{A}_2\cup\mathcal{ A}^{c}}\,,
\end{align}
which is claimed to capture the correlation between $\mathcal{A}_2$ and $\mathcal{A}^{c}$. The evaluation of the mutual information involves the calculation of the entanglement entropy of two disconnected intervals, which is still a formidable task. If we follow the proposal of \cite{Calabrese:2004eu,Hubeny:2007re}
\begin{align}\label{Stwo1}
S_{\mathcal{A}_1\cup\mathcal{A}_3}&=S_{\mathcal{A}_1}+S_{\mathcal{A}_2}+S_{\mathcal{A}_3}+S_{\mathcal{A}}-S_{\mathcal{A}_1\cup\mathcal{A}_2}-S_{\mathcal{A}_2\cup\mathcal{A}_3}\,,
\end{align}
then we find exactly $\mathcal{I}(\mathcal{A}_2,\mathcal{A}^c)=I(\mathcal{A}_2,\mathcal{A}^c)$ which holds for general $\mathcal{A}'$. Though there do exist cases \cite{Casini:2004bw,Casini:2005rm,Casini:2008wt,Casini:2009vk} that apply \eqref{Stwo1}, it is shown in the added note of \cite{Calabrese:2004eu} that the result is in general not correct.

One may also calculate $S_{\mathcal{A}_1\cup\mathcal{A}_3}$ by applying the RT formula to two disconnected intervals as advocated in \cite{Headrick:2010zt} thus the  mutual information \eqref{mutual} is given by
\begin{align}\label{mutualinformation}
I(\mathcal{A}_2,\mathcal{A}^c)=~~\Big\{
\begin{array}{cc}
&\frac{c}{3}\log\frac{l_2 l}{l_1 l_3}\,,~~~~~~~l_2>l_1 l_3/l\,,
\\
&~0\,,~~~~~~~~~~~~~~~~ l_2\leq l_1 l_3/l \,,\\ 
  \end{array} 
\end{align}
which undergoes a phase transition at $l_2=l_1 l_3/l$. When $l_2>l_1 l_3/l$ we have the simple relation 
\begin{align}
e^{\mathcal{I}(\mathcal{A}_2,\mathcal{A}^c)}=e^{I(\mathcal{A}_2,\mathcal{A}^c)}+1\,.
\end{align} 
However this relation does not hold for general $\mathcal{A}'$, because $I(\mathcal{A}_2,\mathcal{A}^c)$ \eqref{mutualinformation} is not invariant under the modular flow.

\bibliographystyle{JHEP}
 \bibliography{lmbib}

\end{document}